# Using Arabic Wordnet for semantic indexation in information retrieval system

**Mohammed Alaeddine Abderrahim[1], Mohammed El Amine Abderrahim[2]**
**Mohammed Amine Chikh[2]**

[1] **Department of Computer Science, University Abou Bekr Belkaid, Faculty of Science, PoBox 119, Tlemcen 13000, Algeria**

[2] **Department of Electrical Engineering and Electronics, University Abou Bekr Belkaid, Faculty of Engineer Science, PoBox 230, Tlemcen 13000, Algeria**

**Abstract**

In the context of arabic Information Retrieval Systems (IRS) guided by arabic ontology and to enable those systems to better respond to user requirements, this paper aims to representing documents and queries by the best concepts extracted from Arabic Wordnet. Identified concepts belonging to Arabic WordNet synsets are extracted from documents and queries, and those having a single sense are expanded. The expanded query is then used by the IRS to retrieve the relevant documents searched. Our experiments are based primarily on a medium size corpus of arabic text. The results obtained shown us that there are a global improvement in the performance of the arabic IRS.

## 1. Introduction

The ontologies are known as tools able to manipulate the knowledge behind the concepts. We can used them in several fields such as informations search, the automatic translation..,. The ontologies can be used at different levels in the IRS. The orjectives of our study is to see the effects of the ontologies in process of indexing documents and queries, we are talking about the semantic indexing. In the literature, there are many definitions of the semantic indexing. The semantic indexation (indexation by the sense of words) aims to correct the problems of the lexical matching by using the semantic indexes rather than the simple keywords. The semantic indexation method aims to retrieve the correct sense of the word in the text from different possibility senses word as defined in dictionaries, ontologies and other language resources [1]. It is based on algorithms of the word sense disambiguation (WSD). Among the disambiguation methods : those combining the disambiguated word with words taken from the context of a document witch help to determine their appropriate sense, more advanced approaches of disambiguation are using hierarchical representation to calculate the semantic distance or the semantic similarity between the compared words[1]. According to Sanderson [2] the successful of disambiguation improves the performance of the IRS, particularly in the case of the short queries (title only). Within the context of using the ontologies for the indexation, we found several works for English language cited in [3], the idea is to built an structure representing the document (respectively query) by using the semantic of the ontologies, this structure is called a semantic core of document (respectively query). Therefore, This is the first work of the semantic indexation of the documents (respectively query) with arabic texts.

In this paper we have implemented the method of semantic indexing of the documents and query for the information retrieval where are use Arabic Wordnet as a semantic resource to exploring the impact of passage from an indexation based on single words to an indexation based on concepts.

This paper is organized as follows. First, we describe the architecture with a discription of the operating process of our system. Then we present the experimentation with a discussion of results achieved and we have finished with a conclusion and prospects.

## 2. Architecture of our System

In this section, we describe the semantic indexing method based on Arabic Wordnet. This approach start with extracting the concepts of wordnet from the documents (respectively query). Then we retrieve the senses of those concepts from the synsets of arabic wordnet and with the method of disambiguation[1] based on calculation of the semantic distances between those senses, we identify the appropriate sense (having only one sense) for every concept from proposed senses. For terms that don't belong to the vocabulary of WordNet, the system extracts their basic form before passing by the semantic indexing method described above. For example, the arabic wordnet does not contain the concept "أسباب", but it contains their basic form "سبب". Formally, let consider: D a document of collection composed of n words.

D= {w1, w2,…, wn}

The result of the concept detection process will be a document Dc. It corresponds to: Dc= {C1, C2,…, Cm, W'1, W'2,…, W'm'}. Where C1, C2,…, Cm are the concepts extracted from the document and identified like wordnet entries. If they are terms that do not belong to the WordNet vocabulary, they are not replaced like the case of words W'1, W '2, ..., W'm'. However, they will be added to complete the representation of the information expressed by the document in order to be used at the search stage.

## 2.1 Details of Our Approach with Example

Let consider the following text of document :

" سواء كانت حالة فقدان الذاكرة بشكل مؤقت أو دائم، أو جاءت بشكل مفاجئ أو ببطء فذلك يعتمد علي أسباب حدوث فقدان الذاكرة. إن عملية تقدم العمر قد ينتج عنها صعوبة في تعلم أو إدراك الأشياء الحديثة علي الشخص أو يمكن أن تتسبب في استغراق وقت أطول من قبل الشخص المسن في تذكر أو استدعاء الأشياء الحديثة عليه (ولكن التقدم في العمر لا يكون سبب في فقدان الذاكرة إلا إذا كان هذا التقدم مصحوباً بمرض معين ساعد في حدوث هذه الحالة). "

Table 1 presents the terms to be indexed after the elimination of the stop words. As well as the segmentation process that is used to link the terms that distinguished only with inflectional mark. Finally, the text is represented by an index of lemmatized words:

Table 1 : List of terms to index

| | | | | | | |
|---|---|---|---|---|---|---|
| فقدان | وقت | مؤقت | دائم | مفاجئ | بطء | يعتمد |
| مرض | ينتج | اشياء | اطول | حديثة | استغراق | سبب |
| جاءت | حدوث | ذاكرة | شخص | تذكر | مصحوب | ساعد |
| تقدم | ادراك | حالة | صعوبة | مسن | استدعاء | عمر |

After omit the stop words, for example: {بعض, سواء}. The process of extracting concepts recognized all the terms of the documents that belong to the Arabic Wordnet. Then, the method of the searching synonyms retrieved all senses of the concepts extracted, and the disambiguation method is used to select the right sense for every concept. The terms that do not belong to the vocabulary of the Arabic Wordnet, they are passed by the module of racine extraction in order to restart the search of the senses with the root. Or else, the words of text will be added to the final index for complete the representation of the information contained in the documents. Table 2 gives an example of selecting indexes to some concepts identified in the text:

Table 2: Example of selecting concepts from Arabic Wordnet

| **Dr/terms** | **Example of Synset Corresponding** | **index choice** |
|---|---|---|
| حدوث | {حَدَث, حُصُول, حُدُوث, ظُهُور, وُقُوع}<br>{حُدُوث, حُصُول, حَادِثَة, حَدَث, وَاقِع} | حُصُول |
| استدعاء | {ذِكْرَى, اِسْتِدْعاء, تَذَكُّر}<br>{اِسْتِدْعاء, طَلَب حُضُور} | تَذَكُّر |
| تذكر | {ذاكِرَة, تَذَكُّر}<br>{ذِكْرَى, اِسْتِدْعاء, تَذَكُّر} | ذَاكِرَة |
| جاء | {أَتَى, جَاءَ}<br>{جَاءَ, ظَهَرَ}<br>{أتى,حَضَرَ,جَاءَ,قَدِم} | أَتَى |
| ذاكرة | {ذَاكِرَة, فِكْر}<br>{ذَاكِرَة,تَذَكُّر} | تَذَكُّر |

For search step, the user queries are expanded with the same method as the documents using the synonyms of those terms to retrieve more relevant results and reduce the silence. Table 3 shows examples of queries before and after semantic indexing method.

Table 3: Examples of queries Expanded

| **N° query** | **Query** | **Proximate concepts** |
|---|---|---|
| 1 | إِثْم | خَطِيئَة |
| …… | …… | …… |
| 12 | بَحْث | دراسة |
| 13 | إِسْتِخْدَام | اِسْتِعْمَال |
| 18 | اِسْتِثْمار | تَوْظِيف |

The detailed of our system are discribed with figure 1:

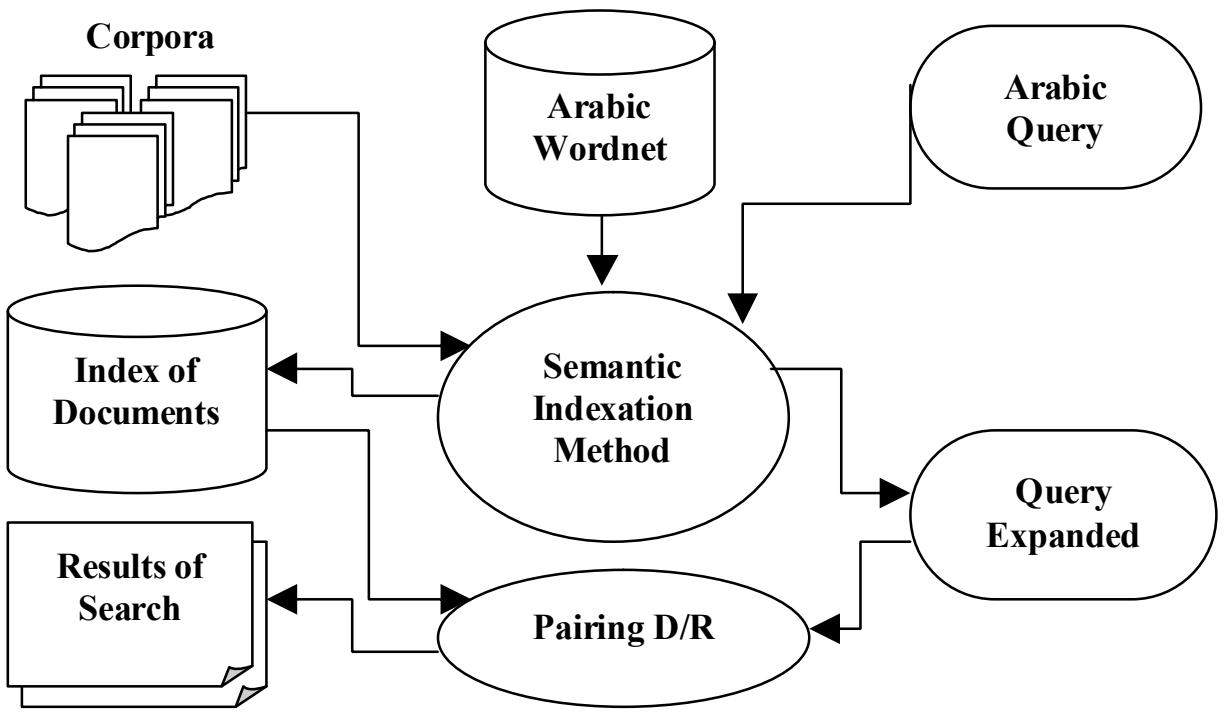

Fig. 1 Architecture of our system

[1] this method choose the appropriate sense (concept) from the proposed senses witch has most linked with other concepts of the same document, the similarity is calculated between senses that belongs to the different sets (synsets).

In the following, we have described our experimentation and discussion the results obtained.

## 3. Description of the experimentations

For our experimentation we have used a corpora of over 22,000 arabic documents (approximately 180 MB) in different areas (health, sport, politic, science, religion, ...). This corpora has approximately 17 millions words with 612,650 are differents word. A set of 70 keywords queries in various fields are chosen for evaluation.

Arabic Wordnet is a lexical database free available for standard arabic. This database follows the conception and methodology of Princeton Wordnet for English and Euro-WordNet for European languages. Its structure is like a thesaurus, it is organized around the structure of synsets, that is to say, sets of synonyms and pointers describing relations to other synsets. Each word can belong to one or more synsets, and one or more categories of discourse. These categories are organized in four classes: noun, verb, adjective and adverb. Arabic WordNet is a lexical network whose nodes are synsets and relations between synsets are the arcs. it currently counts 11,269 synsets (7,960 names, 2,538 verbs, adjectives, 110 adverbs 661), and 23,481 words [4], [5], [6], [7].

To evaluate the semantic indexing method we have segmented our experimentation to four search types and we will study them individually in order to estimate the augmentation of each type to improving the search performance.

The types of search are cited below:

- Simple search or research before semantic indexing (R0): we have used a list of 70 simple queries like keywords with a simple indexation of documents.
- Total Semantic Search (R1): we have indexed semantically a list of 70 queries and the collection of documents used for search.
- Expansion of query (R2): we have indexed semantically only a list of 70 queries and we have used a single word to index the documents.
- Semantic representation of the documents (R3): we have indexed semantically only the database of the documents and we have used a list of 70 simple queries like keywords.

The tables above describe search results:

- The number of documents found.
- The number of relevant documents found.
- The precision at 5 documents (P @ 5).
- The precision at 10 documents (P @ 10).
- The precision at 20 documents (P @ 20).
- The precision at 100 documents (P @ 100).
- The precision at 1000 documents (P @ 1000).
- The median average precision.

Table 4 presents : the number of documents found and the number of relevant documents found.

Table 4: The documents found and the relevant documents for each type of indexation

| N° query | before semantic indexation | | After semantic indexation | | | | | |
|---|---|---|---|---|---|---|---|---|
| | R0 | | R1 | | R2 | | R3 | |
| | Nb Doc Founds | Nb Doc Relevants | Nb Doc Founds | Nb Doc Relevants | Nb Doc Founds | Nb Doc Relevants | Nb Doc Founds | Nb Doc Relevants |
| 1 | 405 | 164 | 11588 | 6287 | 518 | 329 | 8937 | 6092 |
| 2 | 674 | 272 | 9332 | 5071 | 2579 | 1630 | 1914 | 1265 |
| 3 | 366 | 96 | 4237 | 2225 | 3560 | 2163 | 357 | 95 |
| 4 | 3539 | 361 | 17687 | 10985 | 9825 | 5564 | 3781 | 2438 |
| … | … | … | … | … | … | … | … | … |
| 49 | 681 | 423 | 6652 | 3161 | 4860 | 1414 | 663 | 423 |
| 50 | 1578 | 1129 | 6163 | 5267 | 1938 | 1154 | 3077 | 1451 |
| … | … | … | … | … | … | … | … | … |
| 70 | 170 | 50 | 7176 | 3071 | 573 | 297 | 155 | 49 |

A simple comparison of the results obtained before and after using the semantic indexation method to representing the documents and queries, enables us to deduce that this method (for any types) improves in most cases the number of documents and the number of relevant documents returned. In other words, semantic indexing can improve the recall.

Concretely:

- NDTB = The number of documents found before the semantic indexing method.
- NDTA = The number of documents found after the semantic indexing method.
- D = NDTA - NDTB (1)
- NDTPB = The number of Relevant Documents found before the semantic indexing method.

- NDTPA = The number of Relevant Documents found after the semantic indexing method.
- DP = NDTPA - NDTPB (2)
- If (D> 0 or DP> 0) then we can say that semantic indexation improves the performance of IRS in terms of recall.
- In contrast, if (D = 0 or DP = 0), in other words we have the same number of documents returned after the semantic indexing. So, we can say that there are no improvements in the quality of IRS of a recall viewpoint.

Counting The number of queries in terms of D and DP enabled us to establish the results (see table 5):

Table 5: Contribution of semantic indexing based on the documents found and the relevant documents found

| **Documents Found** | | | | | | |
|---|---|---|---|---|---|---|
| | Total queries (R1) | | Total queries (R2) | | Total queries (R3) | |
| D<0 | 0 | 0% | 0 | 0% | 35 | 50% |
| D=0 | 0 | 0% | 9 | 12.85% | 0 | 0% |
| D>0 | 70 | 100% | 61 | 87.15% | 35 | 50% |
| **Relevant Documents Found** | | | | | | |
| | Total queries (R1) | | Total queries (R2) | | Total queries (R3) | |
| DP<0 | 0 | 0% | 2 | 2.85% | 10 | 14.29% |
| DP=0 | 0 | 0% | 4 | 5.72% | 9 | 12.85% |
| DP>0 | 70 | 100% | 64 | 91.43% | 51 | 72.86% |

As shown on table 5, we notice that increasing the number of documents and the relevant documents found covers pratically all queries in R1. Moreover, R2 and R3 are the less appropriate methods for semantic indexing (D <0) and (DP <0) because the use of semantic indexation method modify the vocabulary in documents only (R3) or the queries only (R2). For Example: the term « إثم » it replaced in the semantic index of corpora by « خطيئة » and if we search by using this term query « إثم », the result will be negative.

Based on Table 4, we have established a comparison between the three search types (R1, R2 and R3) in order to identify the best method of semantic indexing of a viewpoint the documents found and the relevant documents found. Table 6 presents the results of this comparaison.

Table 6: Comparison between the various search types (R1, R2 and R3)

| **Documents found** | | | |
|---|---|---|---|
| Percentage of queries which R1 has sent more documents than the others systems | Percentage of queries which R2 has sent more documents than the others systems | Percentage of queries which R3 has sent more documents than the others systems | Percentage of queries which the three systems (R1, R2, R3) sent the same number of documents |
| 85.71% | 4.29% | 0% | 0% |
| **Relevant documents found** | | | |
| Percentage of queries which R1 has sent more relevant documents than the others systems | Percentage of queries which R2 has sent more relevant documents than the others systems | Percentage of queries which R3 has sent more relevant documents than the others systems | Percentage of queries which the three systems (R1, R2, R3) sent the same number of relevant documents |
| 90% | 1.43% | 0% | 0% |

The results described in Table 6 preferred the system R1 so we can say that the semantic indexing of documents and queries together present the best system of search of a viewpoint the number of documents found and number of relevant documents found. This result affirms first consequent which was given in the table (5). Table 7 describes the different values of precision obtained in both systems before and after the use of the semantic indexing method.

Table 7: Different precision values obtained by both systems

| | | **Median Average Precision** | **P@5** | **P@10** | **P@20** | **P@100** | **P@1000** |
|---|---|---|---|---|---|---|---|
| **Before Semantic Indexation** | **R0** | 0,398 | 0,580 | 0,584 | 0,564 | 0,552 | 0,369 |
| **After Semantic Indexation** | **R1** | 0,606 | 0,731 | 0,717 | 0,718 | 0,679 | 0,478 |
| | **R2** | 0,564 | 0,622 | 0,620 | 0,606 | 0,558 | 0,397 |
| | **R3** | 0,551 | 0,600 | 0,620 | 0,602 | 0,579 | 0,389 |

The comparison of three experimentations using the following graphic (see Figure 2) showed us that the semantic indexing method of documents and queries together (R1) give the best rate of precisions in all the measures taken into accounts (P@5, P@10, P@20, P@100, P@1000) as well as the median average precision. whereas, the semantic indexing of documents and queries separately (R2, R3) give inappropriate results for all the measures considered.

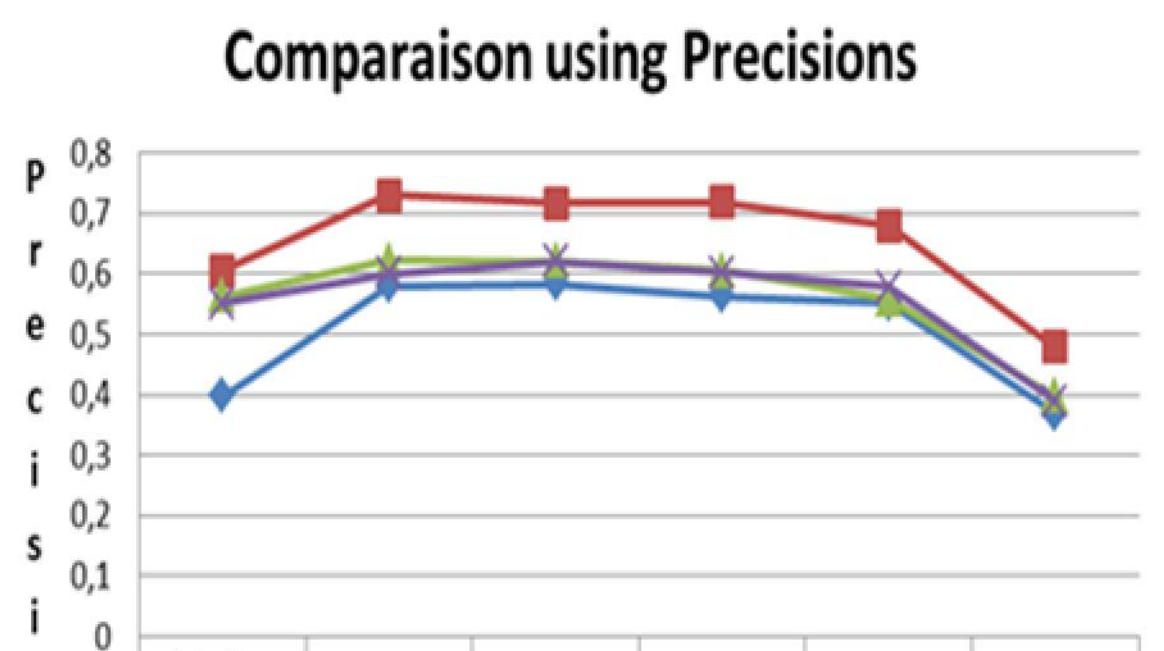

| | Median Average Precision | P@5 | P@10 | P@20 | P@100 | P@1000 |
|---|---|---|---|---|---|---|
| R0 | 0,398 | 0,58 | 0,584 | 0,564 | 0,552 | 0,369 |
| R1 | 0,606 | 0,731 | 0,717 | 0,718 | 0,679 | 0,478 |
| R2 | 0,564 | 0,622 | 0,62 | 0,606 | 0,558 | 0,397 |
| R3 | 0,551 | 0,6 | 0,62 | 0,602 | 0,579 | 0,389 |

Fig. 2 Comparison of precision obtained by different systems

## 4. Discussions

In these experimentation we were interested by testing the semantic indexing strategy to represent the documents and the queries, the implementation of our system is organized as follows: we have started with indexing semantically the collection of documents which is considered as a preparation step for search, by using a semantic resource (as Arabic Wordnet). Then, we have tested different methods of searches started with (R1) which is based on the semantic indexing of documents and queries together. Another way to search, is to index semantically the queries (R2) or choosen to index semantically the collection of documents and use a simple query for search (R3). The objective behind the study of all these methods (R1, R2, R3) is to determine at what level in the IRS, the use of the semantic (in indexation of documents or queries, or together) produces best results.

From the viewpoint documents found and relevant documents found we can say that the use of semantic indexing method to represent both documents and queries together improves the performance of IRS. From the precision viewpoint, (R1) has good values for all the measures considered, consequently, it can be chosen as a method to represent (indexing) information in IRS.

If we must classified the other methods (R2) and (R3), we can said that R2 has the advantage to be more precise for 5 and 10 and 20 and 1000 firsts documents, and the median average precision. Contrary, it presents low values for 100 firsts documents as compared to R3.

The evaluation of the contribution of the arabic ontologies to IRS deduced by this experiment confirms the following characteristics:

- Reducing the silence in response of user queries.
- reduce the noise from responses of queries.
- facilitate the expression of query (assistance in the reformulation of query).
- Increasing the recall and precision.

In this context, we must emphasize that using concepts in the place of terms allows of:

- Provide a good representation of document collections by exploiting the semantics of concepts.
- Facilitating the reformulation of the user query.
- Provide a real support for matching process query/ documents by exploiting the semantic distance existing between the concepts.

## 5. Conclusion

In this paper we have developed an approach that have been proved its force for the English language. The idea of this article is to exploit a lexical resource (Arabic Wordnet) to index the documents as well as the user query in order to improve the retrieval results. Our experiments based on a medium corpus of Arabic language, we have proved that semantic resources (in our case: Arabic Wordnet) improves the quality of IRS and achieving our aims fixed at the beginning. We have remarked that the use of semantic indexing method to represent the documents and the queries together gives better results than using separately. The contribution of the ontologies in information retrieval system with arabic language is very interesting but it requires complete lexical resources witch are not available at present.

It therefore remains many things to do in the future, and the the most imminent extension of our research is to built a semantic core to represent the documents using Arabic Wordnet, as well as the study of the effect of every semantic relationship used in this process like (synonymy, hyponymy,…).

**Mohammed Alaeddine Abderrahim** is a research teacher at the University of Tiaret, Algeria. His research interests are natural language processing, information retrieval, information extraction, data mining and ontology. Med Alaeddine has a Magister in computer science from the University of Tlemcen, Algeria. He is a Doctorate candidate and a member of the Laboratory of Arabic Natural Language Processing in the University of Tlemcen.

**Mohammed El Amine Abderrahim** is a research teacher at the University of Tlemcen, Algeria. His research interests are natural language processing, information retrieval, information extraction, databases and data mining. Med El Amine has a Magister in computer science from UST Oran, Algeria, and a Doctorate in computer science from the University of Tlemcen, Algeria. He is a member of the Laboratory of Arabic Natural Language Processing, university of Tlemcen.

**Mohammed Amine CHIKH** received his Ph.D from the University of Tlemcen, Algeria and is currently a professor at the University of Tlemcen, Faculty of Technologie. His research interests include: knowledge engineering, knowledge discovery, data mining and medical semantic Web. Prof. Med Amine is a project manager in the Laboratory of Génie BioMédical (GBM), university of Tlemcen.